\documentclass[conference,a4paper]{IEEEtran}

\ifCLASSINFOpdf
  \usepackage[pdftex]{graphicx}     \graphicspath{{./figs/}}
      \DeclareGraphicsExtensions{.pdf,.jpeg,.png}
\else
                  \fi

\usepackage[cmex10,fleqn]{amsmath}
\usepackage[fleqn]{mathtools}
\usepackage{amssymb}
\usepackage{pifont}

\usepackage{algorithm}
\usepackage{algpseudocode}

\usepackage{array}

\usepackage[labeled,resetlabels]{multibib}
\newcites{S,P}{Supplementary References,	Postscript References}

\usepackage{float} 
\usepackage{ulem}
\usepackage{color}
\usepackage{soul}

\usepackage{fancyhdr}

\usepackage{colortbl}

\usepackage{afterpage} 
\usepackage{fixltx2e} 
\usepackage{multirow}

\usepackage{randtext}
\catcode`\_=11\relax
\newcommand\email[1]{\_email #1\q_nil}
\def\_email#1@#2\q_nil{  \href{mailto:#1@#2}{{\randomize{#1}\emailampersat \randomize{#2}}}}

\newcommand\emailampersat{{\small@}} \catcode`\_=8\relax

\usepackage{doi}

\usepackage{xfrac}

\usepackage{bm}
\iftrue
\usepackage{url}
\else
\usepackage[bookmarks=false]{breakurl} \usepackage[usenames,dvipsnames]{xcolor}
\hypersetup{
    colorlinks,
    linkcolor={red!50!black},
    citecolor={blue!50!black},
    urlcolor={blue!80!black}
}
\fi

\usepackage{textcase}

\usepackage[font=small,labelfont=bf]{caption}

\usepackage[para]{footmisc}

\begingroup \makeatletter
\gdef\eqna@origamp{&} \catcode`\&\active \gdef\eqna@newamp{  \ifx\@currenvir\eqna@currenvir     \eqna@onlyfirstamp\let\eqna@onlyfirstamp\@empty   \else     \eqna@origamp   \fi
}
\gdef\eqna@hook{  \let\eqna@currenvir\@currenvir   \catcode`\&\active   \let&\eqna@newamp   \let\eqna@onlyfirstamp\eqna@origamp   }
\catcode`\*11 \gdef\eqnarray{\eqna@hook\align} \gdef\eqnarray*{\eqna@hook\align*} \global\let\endeqnarray\endalign
\global\let\endeqnarray*\endalign*
\endgroup

\usepackage{tikz}
\usetikzlibrary{calc,fit,arrows,shapes.misc}

\usepackage{spreadtab}
\STautoround*{2} 
\usepackage{numprint}
\npthousandsep{,}\npthousandthpartsep{}\npdecimalsign{.}

\def\por1{\partial}

\usepackage[binary-units=true,
prefixes-as-symbols=false,
]{siunitx}
\DeclareSIUnit{\microsecond}{\SIUnitSymbolMicro s} 
\newcolumntype{H}{>{\setbox0=\hbox\bgroup}c<{\egroup}@{}}
\newcolumntype{M}{>{\centering\arraybackslash}m{\dimexpr0.25\linewidth-2\tabcolsep}} \newcolumntype{N}{>{\centering\arraybackslash}m{\dimexpr0.10\linewidth-2\tabcolsep}}
\newcolumntype{Y}{>{\raggedleft\arraybackslash}X}

\usepackage{caption}
\DeclareCaptionFormat{myformat}{#1#2#3\hrulefill}
\begin{document}
\bstctlcite{IEEEexample:BSTcontrol} 
\title{Transparent Clouds: An Enhancement to Abstraction}  
\author{\IEEEauthorblockN{\small Reza \MakeTextUppercase{Farrahi Moghaddam}$^{1,2}$}
\IEEEauthorblockA{$^{1}$Synchromedia Lab and CIRROD}
\IEEEauthorblockA{ETS (University of Quebec)}
\IEEEauthorblockA{Montreal, QC, Canada H3C 1K3} 
\IEEEauthorblockA{Email: \email{imriss@ieee.org}} \IEEEauthorblockA{LinkedIn: \url{https://www.linkedin.com/in/rezafm}} \and
\IEEEauthorblockN{\small  Yves \MakeTextUppercase{Lemieux}$^{2}$}
\IEEEauthorblockA{$^{2}$Ericsson Research - Cloud Technology}
\IEEEauthorblockA{Ericsson Canada Inc}
\IEEEauthorblockA{Montreal, QC, Canada H4P 2N2} \and
\IEEEauthorblockN{\small  Mohamed \MakeTextUppercase{Cheriet}$^{3}$}
\IEEEauthorblockA{$^{3}$Synchromedia Lab and CIRROD}
\IEEEauthorblockA{Production Automation Department}
\IEEEauthorblockA{ETS (University of Quebec)}
\IEEEauthorblockA{Montreal, QC, Canada H3C 1K3} 
}

\maketitle

\begin{abstract}
With the introduction of various hardware/software technologies such as Cloud Technologies or Virtualization technologies, there has been a great potential to reuse ICT artifacts thanks to Abstraction and also Exchangeability features achieved via these technologies. These technologies also provide various advantages with respect to sustainability including resource consumption reduction (in the use phase only or in the whole life cycle). However, there is an additional but untapped potential associated with the anonymization of resources introduced by both abstraction and exchangeability features. By realizing on this potential, we can improve cloud solutions and reduce their by-product opacity, which usually prevents leveraging on the specialized but tweakable (i.e., nonessential modifications without changing the main function) features of components that are captured in the component models. This is especially a challenge in the case heterogeneous/disaggregated infrastructure where developing models to cover everything is practically impossible. In this work, by leveraging on the concept of pathways, we develop a few mechanisms that enable transparency and therefore tweakability of features even in the presence of abstraction and heterogeneity. In particular, the layered-stack approach to system decomposition is considered because of its role in both software defined networking (SDN) and Network Function Virtualization (NFV) system decompositions. For a concrete example, the case of dynamic frequency scaling of processors is considered and it is shown that the associated consumption could be considerably reduced without requiring additional changes to the middle components. 
\end{abstract}

\IEEEpeerreviewmaketitle

\section{Introduction}
\label{sec_Introduction}
In the context of sustainability, the life cycle of material, products, and also {\em things} in general has been the one of important focus areas \cite{Pfister2014,Malmodin2014,Stiel2014,Arushanyan2014,Schaubroeck2013,Schien2013,Farrahi2016a,Farrahi2014c,Farrahi2014}, especially in terms of reducing resource consumption. In the domain of the Information and Communication Technology (ICT)\footnote{or more practically (E)ICT that also includes the embedded forms} there are various technologies that pave the way for better use of ICT artifacts available in each ICT solutions. Cloud Technology and Virtualization Technology are two of these enabling technologies, and they have been well recognized and used toward sustainability goals in ICT operations \cite{Farrahi2014d}. 

In \ref{sec_Cloud_and_Virtualization}, we will discuss both Cloud and Virtualization technologies in a greater details. Nonetheless, we can mention that there are two features that are the core of these technologies: i) Abstraction and ii) Exchangeability. Similar to many other use cases of these features \cite{Bourouis2015}, they have led to the introduction of opacity in the operations. Therefore, there has been a well-acknowledged compromise in association with the benefits of these features: Reduced adjustability as a result of opacity.

The paper is organized as follows. In Section \ref{sec_Transparent_Cloud}, the problem statement related to heterogeneous and disaggregated infrastructure is discussed. Then, in Section \ref{sec_NorthwiseSouthwise_Pathways_Approach_to_Transparency_and_Visibility}, the proposed Northwise/Southwise Visibility/Transparency Pathway (NSVTP) approach is discussed, and it is followed with some special cases of targeted visibility. The use case of NSVTP in dynamic frequency scaling of processor cores toward reduction of energy consumption, increase of the equipment life time, and ultimately sustainability is provided in Section \ref{sec_Use_Case_of_Dynamic_Frequency_Scheme_for_Processor_Cores_in_a_Layered_Stack}. Finally, the conclusions and the future prospects are presented in Section \ref{sec_conclusions}. Also, in \ref{sec_Cloud_and_Virtualization}, a generic definition of cloud and virtualization technologies is provided. Then, the layered and networked stacks approach to service decomposition is presented in \ref{sec_Layered_Stacks_vs_Networked_Stacks_NFVNative_Architectures}.

\section{Transparent Cloud}
\label{sec_Transparent_Cloud}
Considering the definition of the cloud computing \cite{ITUY3500201408I}, which requires abstraction of components as the sixth key characteristics (Resource pooling), we define the {\em transparent cloud computing} as an enhancement to standard cloud computing in which the components expose their feature/characteristics as much as possible to each other. In many cases, the service providing components in a pair of connected components sits on underlying layers of the cloud stack. Therefore, these components, which we call them southern components, would be required to provide as much as possible information and control of their features to the service receiving component in the pair (the northern components in this analogy). 

In the context of `homogeneous' infrastructure, the abstracted model of a component would cover almost all the information and controls except the name (ID) of the component. The name (ID) of the components are by definition replaced with the virtualized names (IDs). This means that the cloud computing solution built on top of {\em homogeneous infrastructure} are by default transparent. 

\subsection{Heterogeneous Infrastructure and Transparent Cloud}
\label{sec_Heterogeneous_Infrastructure_and_Transparent_Cloud}
In contrast to homogeneous infrastructure, the concept of `transparent cloud' requires special attention in the context of the emerging heterogeneous/disaggregated infrastructure (especially hardware part). With the emergence of many new manufacturing/designing players in the hardware domain, the solutions (such as cloud) have no choice except to operate with a highly `heterogeneous' set of hardware components. Even in the case of only a single manufacture, the set of possible hardware components is growing mainly in response to the diverse compute needs of the new applications. On top of that heterogeneous set of hardware components, the number of the possible ways to combine these hardware components is much bigger. It has been observed that a heterogeneous/disaggregated infrastructure is of great advantages for both the provider and the customer in terms of reducing the total cost and also overprovisioning \cite{Mainstay2016}. 

It has been suggested to use a combination of managers and orchestrators in order to handle and operate a multi-tenant compute center (such as that of the case of virtualized network functions \cite{Bogineni2016}). However, development of a true cloud solution that has the capability to take the maximum advantage from a heterogeneous infrastructure is a challenge mainly because of the abstraction aspect of the cloud solutions. A particular case of this challenge is discussed in the next section.

\subsection{The Challenge of Resource/Component Rotation in Heterogeneous Infrastructure}
\label{sec_The_Challenge_of_ResourceComponent_Rotation_in_Heterogeneous_Infrastructure}
A particular case in this context is the case of component `rotation', which is an advantage of the cloud solution. In a component rotation action, an underlying component (which probably has failed) is replaced with another component with the same features and in the same layer without requiring any changes to the overlying components. This is thanks to the inherent abstraction feature of the cloud solutions. 

However, in the case of heterogeneous infrastructure, the cloud operator may encounter cases in which they cannot find an available, replacing component with the same features, and therefore they are forced to use a component with `higher' grade features in order to complete the rotation action. As mentioned before, avoidance of overprovision is a key advantage of heterogeneous infrastructure, and therefore the cases when a high-end component is used in place of a low-end component would be non-zero. Although the solution will work thanks to abstraction, the extra features of the high-end component will not be used. If we can find a way to expose the extra features of the replacing component to the northern component, not only higher performance could be achieved, the user experience (in this case, the tenant experience) would be increased by becoming aware of the free upgrade they received.

It is worth noting that there are no changes to the main service interactions among the components. In other words, the additional mechanism does not require any modifications on the connections in the cloud stack. Instead, it provides an extra means of communications between two components in order to adjust their states according to the preference of the other component even if that change in the state has not been captured by the abstracted model of the component. An example of such states is the clock frequency of CPUs. Changing the state of a CPU with respect to its clock frequency (using the particular features provided by the manufactures) would not require any change in the main stream of instructions being executed. However, such a change could be of the interest of the application for many reasons including cost saving, footprint reduction, and even demand response.

\begin{figure}[tbh!]
	\centering
	\begin{tabular}{cl}
		(a) & \includegraphics[height=0.8in]{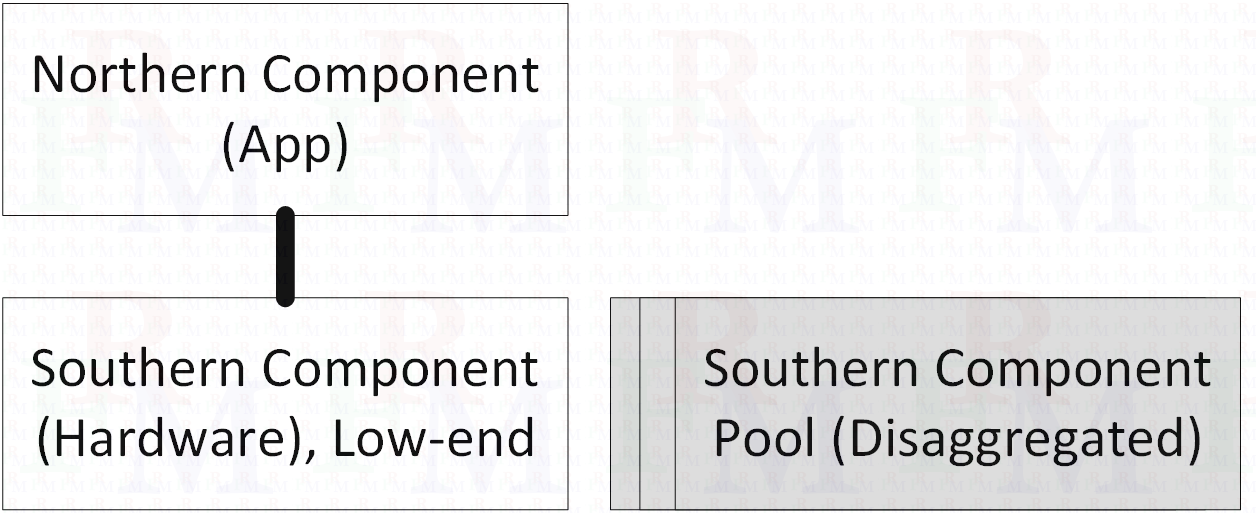}  \\\hline
		(b) & \includegraphics[height=0.8in]{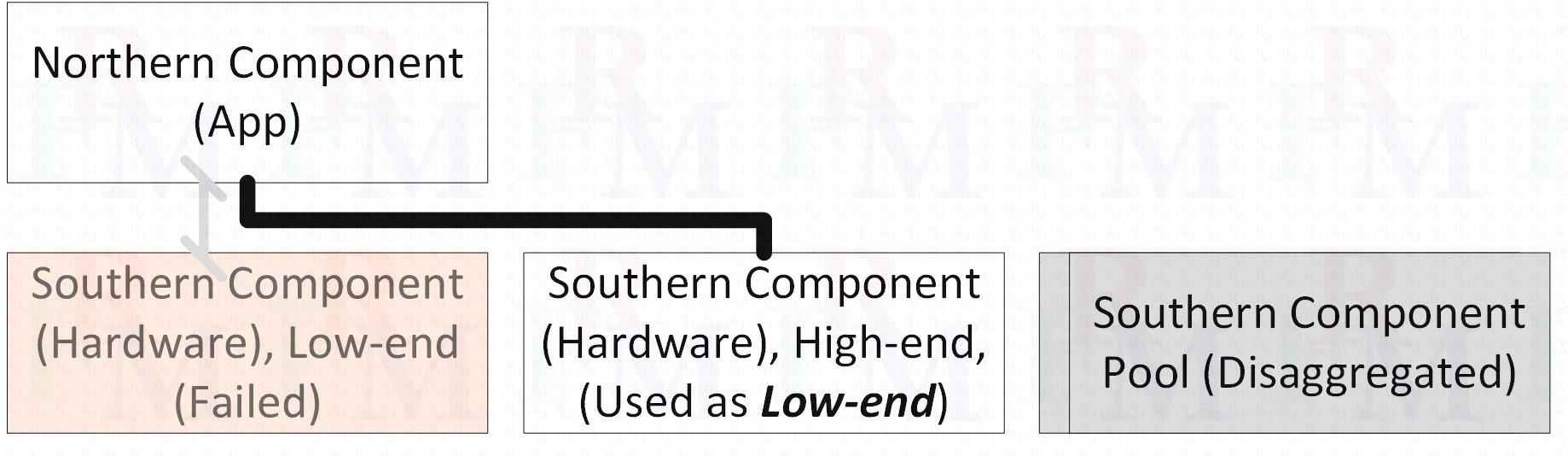}  \\\hline
		(c) & \includegraphics[height=0.8in]{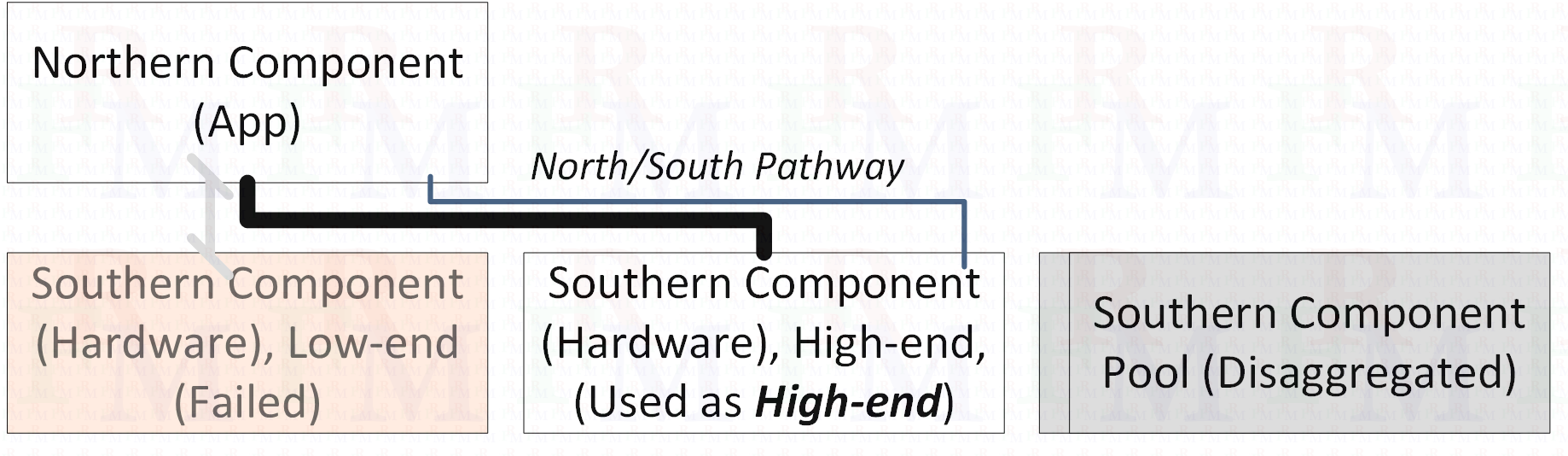}  
	\end{tabular}
	\caption{A typical example of the application of NSVTP in a heterogeneous/disaggregated infrastructure environment.
		a) The requested setup.
		b) The configuration after failure of the southern component. A high-end (not required in the requested setup) is pulled from the pool of disaggregated hardware and used as a low-end component (low utilization).
		c) The same as (b) but with the NSVTP pathways enabling the northern component to adjust/modify the high-end features of the new southern component (high utilization possibly with extra billing). No other modification is needed.}
	\label{fig_NSVTP_Concept1}
\end{figure}

Figure \ref{fig_NSVTP_Concept1} shows a typical use case in which a southern component fails and is replaced by a high-end component. Two alternatives are considered. In the first one, there is no additional mechanism to expose the features of the new southern component to the northern counterpart. In the second alternative, which is the case of this paper, a direct messaging mechanism allows the exposure, and then consequently the northern component can adjust the state of the northern one to its preferred state in order to reduce the cost or improve the performance. In the case of homogeneous hardware, there will be no significant advantage for the second case, and both cases could be considered the same in terms of transparency. 

\subsection{NSVTP Approach to enable Transparent Cloud}
\label{sec_NSVTP_Approach_to_enable_Transparent_Cloud}
Instead of redefining the complete stack of a cloud solution, here, we proposed an alternative that acts as an `connection' (but not within the stack) in order to enable components (hardware, software, and services) to expose their specific features/states to each other (especially those in the higher layers) without requiring any modifications to the `middle' components. This proposed solution is based on an additional obstructed messaging between two components. Therefore, it does not affect the main message/API calls that passes among the components and services in their normal operation. Instead, it allows a component to request another component to change its state in order to better fit with the requirements of the former one. To reduce its impact to as low as possible, the messages are exchanged in the form of component names. In other words, the names are generalized in order to make them capable to carry the exposure and control messages without requiring addition of any extra services. The details of the proposed approach, which is called Northwise/Southwise Visibility/Transparency Pathway (NSVTP), is provided in the next section. It is worth mentioning that the proposed NSVTP approach and its mechanism is not the only way to realize the concept of transparent cloud in the context of heterogeneous infrastructure.

\section{Northwise/Southwise Pathways Approach to Transparency and Visibility}
\label{sec_NorthwiseSouthwise_Pathways_Approach_to_Transparency_and_Visibility}
The purpose of the Northwise/Southwise Visibility/Transparency Pathways (NSVTP) feature is to enable direct, immediate interaction and communication between two components in a stack while having a `minimal' impact or footprint on the functions and APIs of any middle-layer component involved in the operation. To achieve such functionality, we break down the NSVTP in the two pathways: One to carry communications to the `north' and one to handle the communications to the `south'. It is worth mentioning that here we use the notions of north and south with respect to vertical dimension of the stack, and they should not be mistaken with the similar notions used in the context of chipset bridges.

Along each one of the two pathways, i.e., the northwise and southwise pathways, the NSVTP interactions are implemented in the form of some appendices, which we call `capsules', that are added to the component IDs. Considering the fact that almost all currently available stack designs use a name-based approach to enabling interactions among their components, we assume that every component is assigned with a unique component ID (i.e., resource ID). In the following sections, we discuss how the NSVTP approach uses and also extends the name IDs in order to enable minimal-footprint insertion of direct interaction between not-immediately-connected components in a stack. Again, it is worth mentioning that these additional, inserted interactions/communications should not be mistaken with the actual interactions/communications between the components. The actual solution is carried forward using the actual interactions, while the additional NSVTP interactions enable modification of nonessential features of the (southern) components.

\begin{figure}[tbh!]
	\centering
	\begin{tabular}{c}
		\includegraphics[width=3.5in]{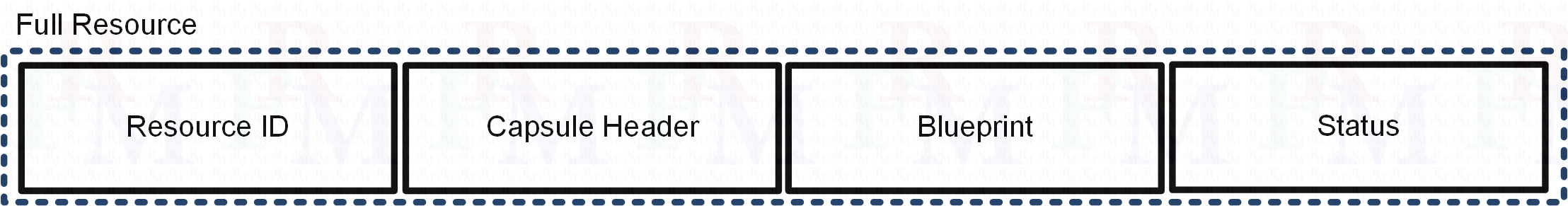} 
	\end{tabular}
	\caption{The structure of a typical extended name ID including the northwise capsule appendix. The subparts of the capsule, i.e., the header, the blueprint, and the status segments, are also shown.}
	\label{fig_NSVTP_Capsule_Northwise1}
\end{figure}

\begin{figure}[tbh!]
	\centering
	\begin{tabular}{c}
		\includegraphics[width=3.5in]{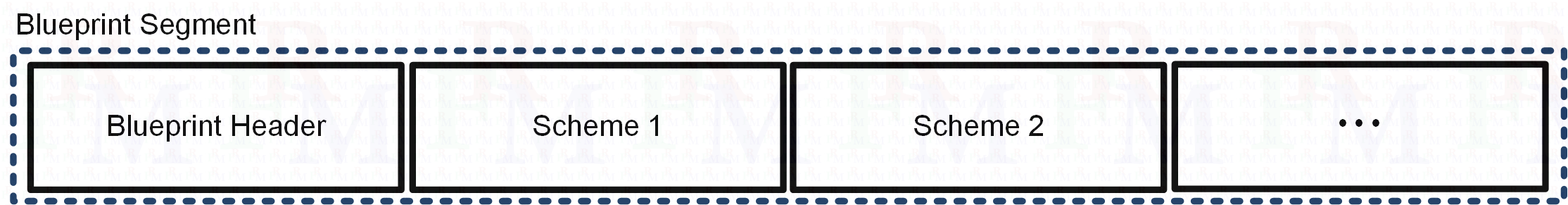} 
	\end{tabular}
	\caption{The structure of a typical blueprint segment of a northwise capsule. A header followed by a plurality of schemes constitute the blueprint segment.}
	\label{fig_NSVTP_Capsule_Northwise_Blueprint1}
\end{figure}

\begin{figure}[tbh!]
	\centering
	\begin{tabular}{c}
		\includegraphics[width=3.5in]{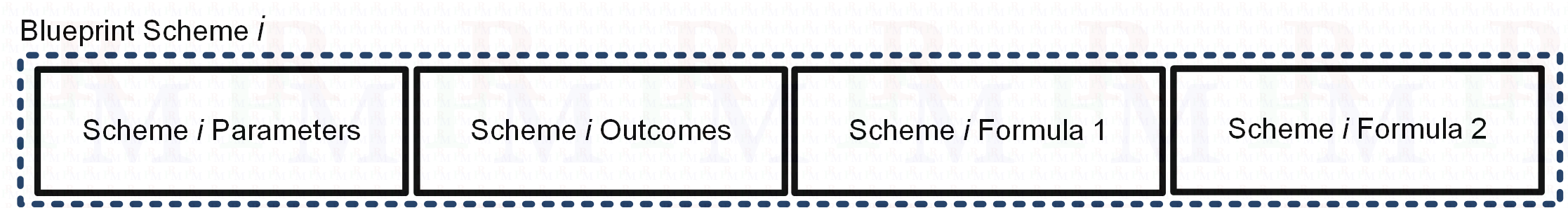} 
	\end{tabular}
	\caption{The structure of a typical blueprint scheme. It consists of the parameters, the outcomes, and a plurality of formula.}
	\label{fig_NSVTP_Capsule_Northwise_Blueprint_Scheme1}
\end{figure}

\subsection{Northwise Blueprint/Status Capsules}
\label{sec_Northwise_Blueprint_Status_Capsules}
The core of the Northwise capsules is an extension to the `component' (`resource') identifier.\footnote{The extension is considered as an option.} The advantage of this form of direct communication and interaction between two components is that there is no additional requirements or functions that the `middle' components would be required to provide. In other words, the two components directly `call' each other, and the middle components just pass the names along the way up (north) or down (south). In the proposed extension to the resource identifier, shown in Figure \ref{fig_NSVTP_Capsule_Northwise1}, a regular resource identifier (called a resource ID from here on) is augmented with an {\em appendix} that carries the capsule data. In many of the NSVTP messages or communications, the appendix could be omitted if the required data has been already communicated between the two components (and has been constant afterward). Figure \ref{fig_NSVTP_Capsule_Northwise1} provides a basic structure of the extended identifier. It starts with the resource ID, i.e., the `regular' resource identifier assigned to the component, and then it carries the capsule header followed by the capsule's actual data. The capsule header is used to differentiate among various types of capsules and also other possible appendices. The capsule data itself is divided into two subparts or segments: i) the blueprint and ii) the status. The blueprint segment is the core of the capsule, as described in Figure \ref{fig_NSVTP_Capsule_Northwise_Blueprint1}, in which a typical blueprint segment is shown. A blueprint consists of a header followed by a plurality of blueprint's {\em schemes}. Each scheme itself, as shown in Figure \ref{fig_NSVTP_Capsule_Northwise_Blueprint_Scheme1}, consists of its associated parameters and {\em formula}. We assume that each scheme could have more than one formula in order to account for various {\em regimes} of operation of components without over-fitting or under-fitting a formula in order to cover all the regimes.\footnote{Every component could have different behaviors when it is subject to different input. For example, a component could have different {\em regimes} of operation at its nominal, over-nominal, or under-nominal input (activity) levels. If a model is forced to cover all these regimes, it is usually prune to under-fitting, i.e., lower but constant accuracy. In contrast, if a very small interval is used in the modeling, the model would be an over-fit, i.e., claiming a high accuracy but with the risk of unbounded errors outside the modeled interval.} 
As an example, the scheme of the processor cores with respect to their dynamic frequency is describe in Figure \ref{fig_UseCase_DVFS1}. 

The status segment of the capsule data is used to transfer or inquire about the status of a southern component or the status of its constituents. The status data is populated using the parameters followed by the values in a simple to interpret way such as that of JSON data structures. In order to be more formalized, a scheme in the blueprint is reserved to formulate and provide the parameters and actions related to status segment.

By default, it is assumed that the capsule segments or the whole capsule itself are encoded (for example compression encoding for a reduced size). This allows for lowering the cost and communication load associated with the capsules transfer and also messaging. The encoding can be also replaced or enhanced with encryption in those cases where the capsule data is assumed to could potentially `interfere' with the operation and performance of the `middle-layer' components. However, it is also optionally possible to skip/omit encoding or encryption of the capsules at all. It is worth mentioning that the whole capsule can be enveloped in a Type-Length-Value (TLV) element as an optional information in order to be transparent to the involved protocols across the stack \cite{Dearlove2015}. Also, there is no fixed space/size reserved for the segments in the capsule structure in order to save space when a segment is constant compared to previous communicated instances of an extended resource ID. When a segment is constant with respect to its previous description, it is replaced with NULL characters from the ``alphabet'' used to describe the resources or capsules. Also, by default, the blueprint and the status segments are encoded/encrypted separately from each other and then are concatenated. However, they could be first concatenated and then are encoded as a single entity, if preferred (or because of possible difference in the encoding performance in the actual setup).

\subsection{Southwise Tweak Capsules}
\label{sec_Southwise_Tweak_Capsules}
In the southwise communications, the capsule structure is similar to that of the northwise communications mentioned in the previous section. The main difference is that the appendix is composed of a capsule header followed with a plurality of {\em tweaks}. A `tweak' is defined as an {\em populated} version of one of the formula from the associated blueprint schemes previously communicated in a northwise capsule to the northern component. By `populated' we mean that the parameters of the scheme are set by the northern component.

\subsection{Trusted eXchange (TX)}
\label{sec_Trusted_eXchange}
In cases where an encoded/encrypted NSVTP interactions is preferred, a trusted component is considered to handle the key exchanges and also verify the layer index of the components. The southern component $s$ encapsulates an instance of its blueprint $B_s$ in a capsule $C_s$ using encoding/encrypting approaches. Then, it again encapsulates the capsule $C_s$ and the associated key in another capsule $C'_s$ probably using another approach (i.e., $k'$). Then, it sends {\em up} the second key $k'$ to middle layers and finally to the {\em targeted} northern component on the $l^\text{th}$ layer, and at the same time sends the tuple $(C'_s, l)$ to the Trusted eXchange (TX), where $l$ represents the layer {\em index} of the northern component. The northern component is assumed to contact the TX with presenting a tuple consisting of i) the key received, ii) its layer index, and iii) a key for southwise capsules ($k''$). We assume that the TX is capable to verify the {\em layer index} of the incoming requests. Then, the TX uses the key provided by the requester, i.e., the northern component, and retrieves $C_s$ and its associated key, and then sends them to the requester. 

In many cases, the TX could be practically omitted. Then, direct exchange of blueprints and tweaks between the southern component and northern component is used. The two parties could still use encoded/encrypted copies in the exchanges in order to reduce the size of messages being exchanged. However, verification of the layer index would be a challenge if required.

\begin{figure}[tbh!]
	\centering
	\begin{tabular}{c}
		\includegraphics[width=3in,trim=0 0 1.75in 0,clip=true]{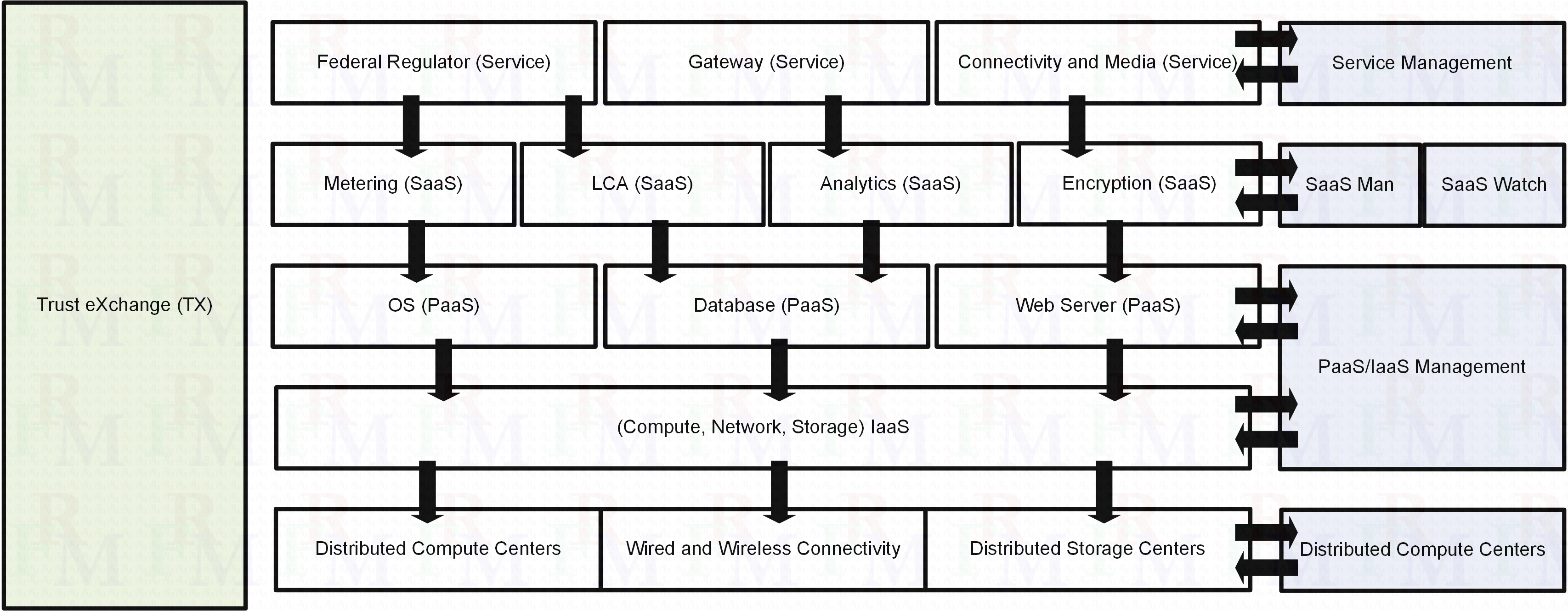} 
	\end{tabular}
	\caption{Layered Service Stack along with its Trust eXchange.}
	\label{fig_schematic_metro1}
\end{figure}

\begin{figure}[tbh!]
	\centering
	\begin{tabular}{cc}
		\fbox{\includegraphics[width=1.5in,trim=0 0 3.5in 0,clip=true]{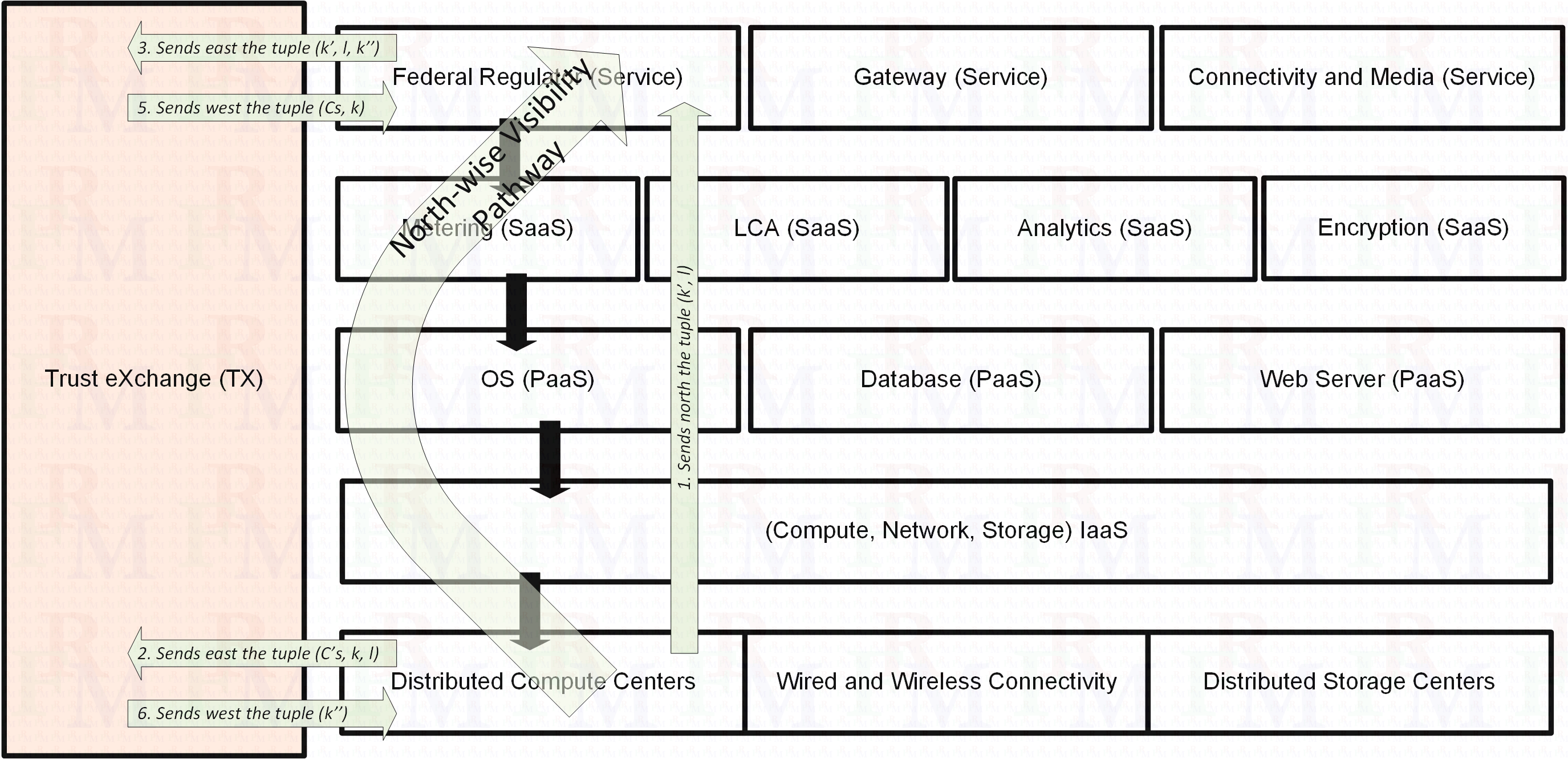}} &
		\fbox{\includegraphics[width=1.5in,trim=0 0 3.5in 0,clip=true]{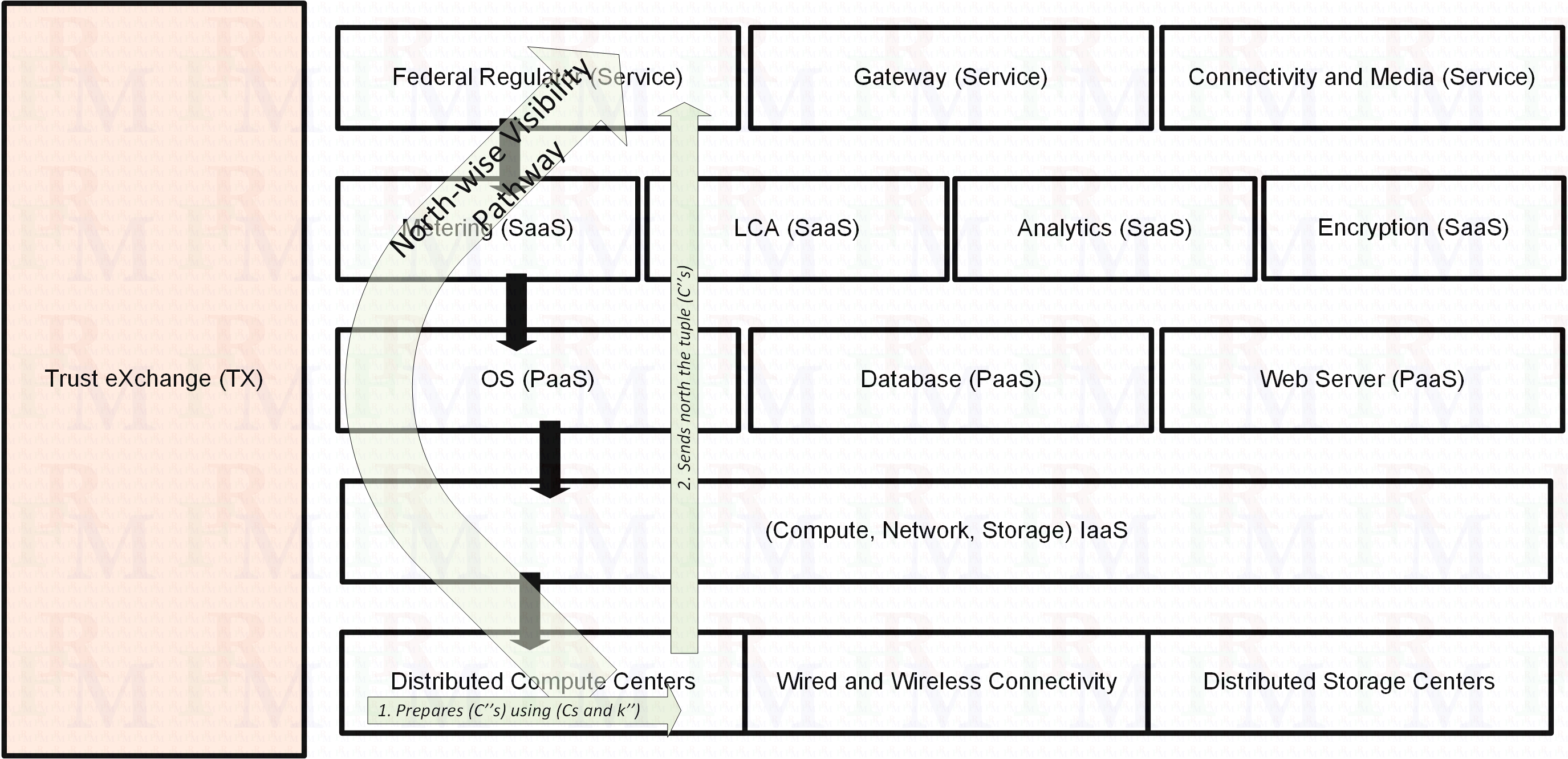}} \\
		(a) & (b)	
	\end{tabular}
	\caption{The time steps of a northwise path establishment using the Trust eXchange in a Layered Service Stack.
		a) The initiation steps.
		b) Northwise communication after the initiation steps.}
	\label{fig_Layered_PlusTX_NorthPath_Implement1}
\end{figure}

\begin{figure}[tbh!]
	\centering
	\begin{tabular}{@{}cc}
		\fbox{\includegraphics[width=1.5in,trim=0 0 3.5in 0,clip=true]{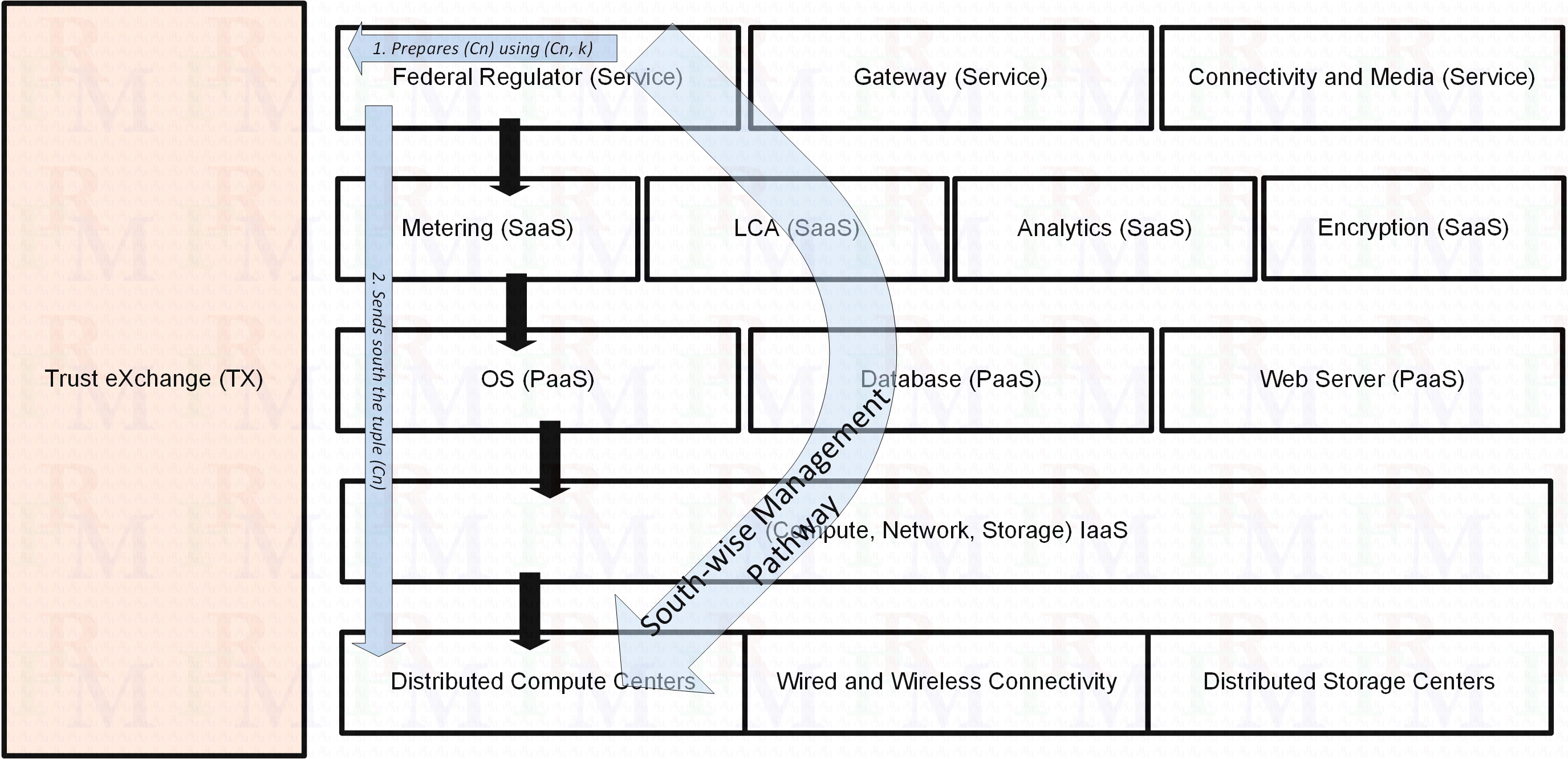}}  &
		\fbox{\includegraphics[width=1.5in,trim=0 0 3.5in 0,clip=true]{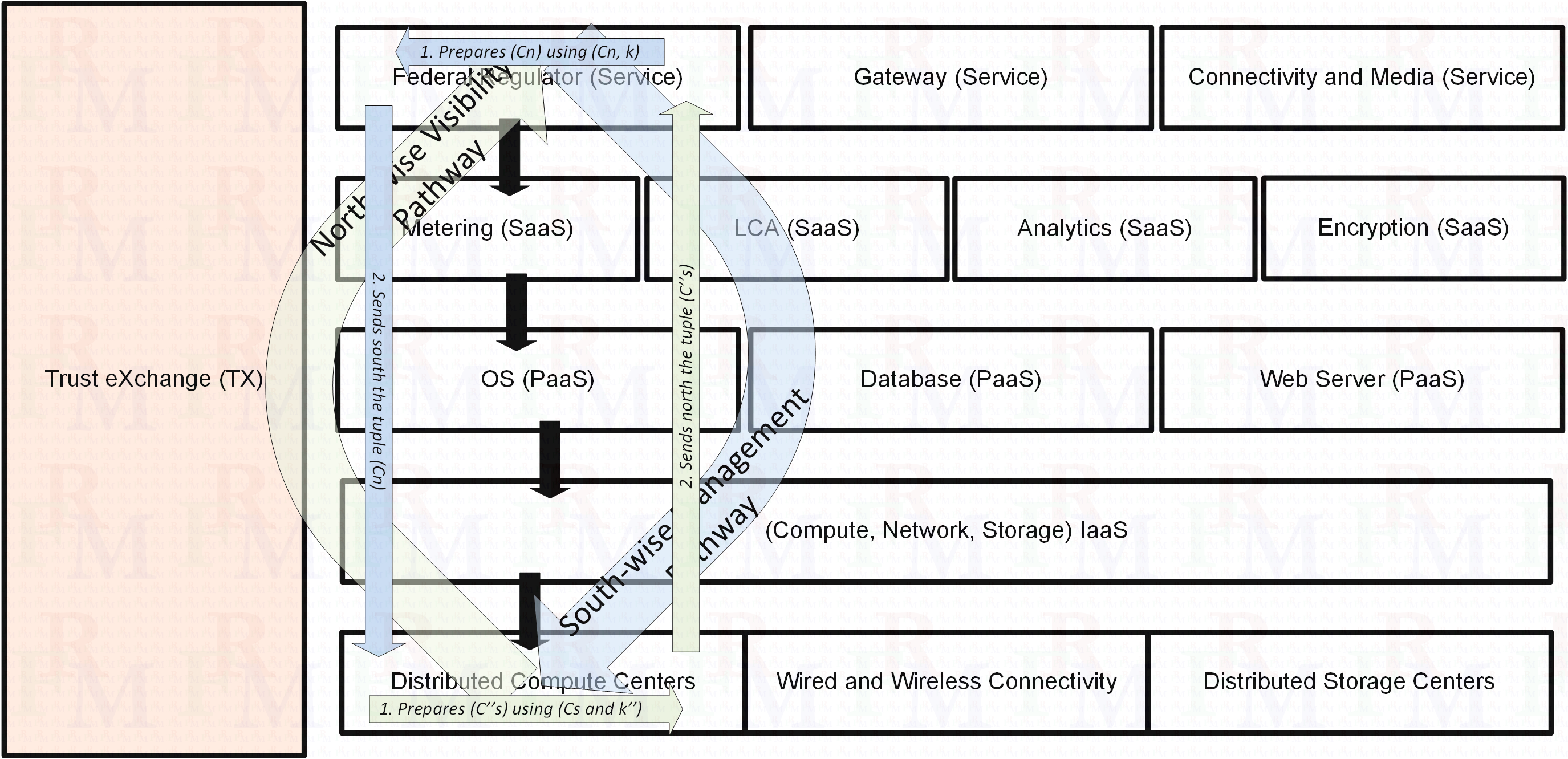}}  \\
			(a) & (b)
	\end{tabular}
	\caption{
		a) The time steps of a southwise path establishment using the Trust eXchange in a Layered Service Stack.
		b) The overall north/south interaction after initiation step (no more need for the TX mediation).}
	\label{fig_Layered_PlusTX_SouthPath_Implement1}
\end{figure}

The actual step to perform and initiate a NSVTP between a southern and a northern components is shown in Figure \ref{fig_Layered_PlusTX_NorthPath_Implement1}(a). After the initiation steps, the southern component could directly communicate and call the northern component, as shown in Figure \ref{fig_Layered_PlusTX_NorthPath_Implement1}(b). At the same time, the northern component directly calls the southern component after the initiation process (Figure \ref{fig_Layered_PlusTX_SouthPath_Implement1}(a)). The complete bi-directional NSVTP interaction between the two components is shown in Figure \ref{fig_Layered_PlusTX_NorthPath_Implement1}(b).

\begin{figure*}[tbh!]
	\centering
	\begin{tabular}{c}
		\includegraphics[width=6in]{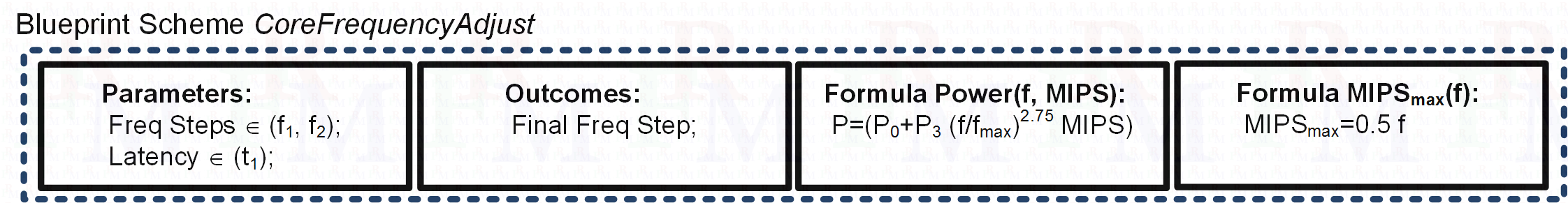} 
	\end{tabular}
	\caption{A Blueprint Scheme related to DVFS use case.}
	\label{fig_UseCase_DVFS1}
\end{figure*}

\section{Use Case of Dynamic Frequency Scheme for Processor Cores in a Layered Stack}
\label{sec_Use_Case_of_Dynamic_Frequency_Scheme_for_Processor_Cores_in_a_Layered_Stack}
Power consumption and performance modeling and management in data centers is of great interest \cite{Piga2013,Bulej2016,Bures2015,Weaver2015,Lopez2015,Wang2016}. The Dynamic Frequency Management of processor/cores' clock frequency has been considered and used as a alternative option to consolidation in saving energy and also in increasing the life time of equipment \cite{Farrahi2014c,Farrahi2014,Mazouz2014}. This approach is usually referred to as Dynamic Voltage and Frequency Scaling (DVFS) or other similar terms that actually refer to those processes that are used to dynamically change the clock frequency. However, from here on, we use the term DVFS in a generic form to cover all realization processes for the purpose of simplicity. It has been shown that DVFS-enabled, workload-aware managers in the data centers and clouds could highly reduce the OPEX of the operation thanks to saving in the energy consumption by the processors \cite{Farrahi2014}. This saving is usually rooted in a `nonlinear' relation between the electricity power requirement of a core and its clock frequency. In its simplistic form, a cubic relation could be inferred. However, in the real cases, the power-law index is usually less than 3:
$$ P=(P_0 + P_3 (f/f_\text{max})^{n_\text{DVFS}} (l/l_\text{max})$$
where $P$ is the power required by the core, $f$ is the current clock frequency, $f_\text{max}$ is the maximum clock frequency possible, $l$ is the load on the core in the Million Instructions per Second (MIPS) unit, and $n_\text{DVFS}$ is the associated power-law index.

Also, it can be argued that many applications, such as Map-Reduce, follow a form of deterministic cycles of `isolated intense compute' followed by `inter-node intense transfer` \cite{Zahavi2014}. Although proposing new patterns for application could be a necessity, optimizing the resource and application based on the current patterns is another requirements for the cloud manager/controller and in general for the Data Center Infrastructure Management (DCIM). In what follows, we provide an example of how NSVTP approach could be used in order to leverage on such patterns in applications.

The adjustment of DVFS by the manager/controller of the cloud system is great. However, it could be `skipped' by the application developers who could see it as a risk factor for their performance. At the same time, the un-guaranteed latency introduced because of the communications required to the manager/controller and then from there to the southern components makes it practically impossible to perform the DVFS adjustments in very `fine' granularities along the time. As mentioned before, there is a great potential for optimization and energy saving in fine granularities because of periodic patterns of compute/communicate cycles in many applications. This brings us to the proposed NSVTP approach which enables real-time, direct communication between a northern component (probably an application) and a southern component (such as a CPU core) in order to fully utilize the benefits of dynamic adjustments in the clock frequency.

Figure \ref{fig_UseCase_DVFS1} shows an example of blueprint scheme that could be probably provided by a southern CPU core to a northern component in order to set up a DVFS `tweak'. There are two parameters, the frequency step and the latency. Each parameter has a set of feasible values to select from. For example, the latency parameter has only one value ($t_1$) in its feasible set which means each DVFS interactions from the northern component would take at least a time interval of $t_1$ in order to change the frequency. That means if the compute/commute (or intense/relax) pattern of the application has a time period around or smaller than $t_1$, the dynamic frequency scaling would not have any benefit to the operation. This shows the capability of the NSVTP in enabling DVFS at the time scales as small as the latency limits of the DVFS changes. This potential can be leveraged toward energy saving, increase in the performance, and in general sustainability. Considering the norm that the DCIMs usually modify the state in a long period of time (such as one hour), it is clear that the NSVTP operating at short time intervals (such as one second or less) could bring in semi-autonomous, decentralized management mechanisms without requiring any modification neither in the middle-layer components nor in DCIM components. In the following subsection, we provide some estimations of the impact of NSVTP addition to a cloud computing operation.

The latency in the dynamic frequency scaling interactions could range up to $20$~\si{\microsecond} \cite{Suh2009}. However, it should be pointed out that this time interval could be actually a halt/pause in the processor/core operation during which the voltage and frequency is adjusted. In this work, we assume that this time interval is only a transition time interval in which the processor/core's circuit is active but not stable (such as that of the case of Marvell XScale processors \cite{Poole2009}).

\subsection{Use Case of Dynamic Frequency Scheme: Experiments}
\label{sec_Use_Case_of_Dynamic_Frequency_Scheme_Experiments}
Although each application could have a different configuration requirements or best practices, we assume that application of interest in our study of long-term with periodical compute-commute intervals. Also, we assume that the applications are hosted in a cloud-oriented data center with disaggregated hardware at the cabinet (rack) scale \cite{Ericsson2015e,Mainstay2016}. That means that the hardware component available in a cabinet could be automatically mixed-and-matched in timely manner in order to build desired compute nodes up to the maximum capacity of the cabinet. It is worth mentioning that, even without NSVTP, disaggregated hardware provides a considerable amount of energy saving and reduction in footprint thanks to its ability to reduce over-provisioning and to ``recycle'' the hardware components.

In the experiments, we use the same specifications for the DVFS feature as those of \cite{Farrahi2014c}: $P_0=142.2$W and $P_3=107.8$W,  $f_\text{min}=1$~GHz, and $f_\text{max}=3$~GHz. Also, the performance of the hardware (thanks to disaggregated granularity) is assumed to be $80$\%, i.e., on average for a workload of intensity $1$ MIPS, we would allocate a $1.2$ MIPS core equivalent because of discrete nature fo the number of cores. This overhead is because of discrete nature of the number of cores, and it is not specifically dependent on $1$~MIPS. For example, for cores of a power of $3$~MIPS, all workloads with a MIPS less than $3$~MIPS would be assigned at least one core. Resource sharing practices would allow other workloads to use the extra compute power of a core. However, because of requirements such as isolation, limiting management overhead per resource, and handling unpredicted workload variations, the maximum performance would be limited below 80\%.

In the first case, we assume only one application. Considering the disaggregated nature of the data center, this means that we are ignoring the rest of application in this case, and therefore there is no conflict with multi-tenancy operation of the data center. The cycles of compute-commute of an application $i$ can be characterized by the ratio compute sub-interval to the data exchange sub-interval ($\rho_i$) and also the compute sub-interval itself ($t_{\text{comp}, i}$). These two application characteristics could be modeled using statistical distributions. However, here we assume that they are represented by their mean values. 
The performance of the application without using NSVTP in terms of energy consumption could be calculated as follows: Over each cycle of compute-commute pattern, the associated processor/core would consume $\left(P_0+P_3\right)\left(t_{\text{comp}}\right)\left(1+1/\rho\right)$. When the NSVTP is used, assuming $\delta<2*t_{\text{comp}}/\rho$, the consumption could be calculated as follows: $\left(P_0+P_3\right)\left(t_{\text{comp}}+2\delta\right)+\left(P_0+P_3\left(f_\text{min}/f_\text{max}\right)^3\right)\left(t_{\text{comp}}/\rho-2\delta\right)$. Therefore, the overall improvement because of NSVTP would be:
\begin{eqnarray}
\eta & = & \frac{\left(1+2\delta/t_{\text{comp}}\right)}{\left(1+1/\rho\right)} \nonumber \\
&& +\frac{\left(P_0+P_3\left(f_\text{min}/f_\text{max}\right)^3\right)}{\left(P_0+P_3\right)}
\frac{\left(1/\rho-2\delta/t_{\text{comp}}\right)}{\left(1+1/\rho\right)}
\end{eqnarray} 

\begin{figure}[tbh!]
	\centering
	\begin{tabular}{c}
		\includegraphics[width=3.5in]{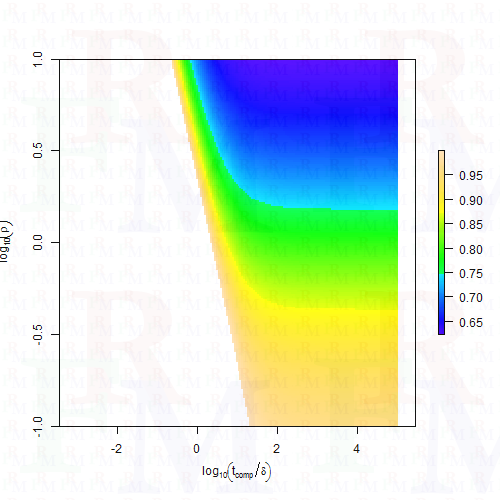}
	\end{tabular}
	\caption{The performance of NSVTP in the case of a single application. The performance is plotted versus two dimensions: The relative $t_\text{comp}/\delta$ and $\rho$. A performance value of for example 0.7 in the figure would indicate a 70\% decrease in energy consumption compared to the baseline. The minimum value for the performance is around 60\% in this case.}
	\label{fig_UseCase_DVFS_SingleApp1}
\end{figure}

Figure \ref{fig_UseCase_DVFS_SingleApp1} shows the impact of these parameters on the performance of the NSVTP approach in reducing the energy consumption of the application in the compute nodes. Because we do not optimize the behavior of the application in this work, the energy consumption and other impacts of the exchange and transfer of data sub-intervals would be the same (therefore, we ignore the modulation phenomena). As can be seen from the figure, the NSVTP mechanism provides an additional minimum of 60\% relative improvement compared to the baseline operation, i.e., not using the NSVTP mechanism. The minimum improvement is related to high $\rho$ value regions of operation that are related to those applications that have long period of compute-intense time intervals compared to their data-exchange time intervals. It is worth mentioning that the key factor in `lower' performance (but still positive) of the NSVTP mechanism for high-$\rho$ regions is the presence of disaggregated hardware in our use case. In other words, the disaggregated hardware feature (along with presence of a heterogeneous hardware resource pool) allows allocation of almost-optimal resources to applications with and without inclusion of the NSVTP mechanism. Therefore, for settings that are not using the disaggregated hardware feature in their cloud solution, the positive impact of the NSVTP mechanism would be much higher. We consider such cases in a future work.
 
\subsection{Impact of N/S Pathways in Reduction of Footprint Associated with Widespread Adaptation of Cloud Computing and ICT}
\label{sec_Impact_of_NS_Pathways_in_Reduction_of_Footprint_Associated_with_Widespread_Adaptation_of_Cloud_Computing_and_ICT}
Considering the forecasts that the data center market will grow rapidly in the coming years (such as 22\% Compound Annual Growth Rate (CAGR) forecast for the Nordic region market \cite{WGR2016}), there will be a considerable potential for the NSVTP-based approaches and mechanisms to reduce the energy consumption of the data center and cloud operations which in turn improve the performance and also reduce the associated environmental footprint. With the minimum of 60\% improvement obtained for the DVFS use case in the previous section, combined with the disaggregated hardware performance, the NSVTP approaches and other similar decentralized management and operation mechanisms would bring a considerable level of energy and cost saving, which in turn translate in lesser environmental footprint and higher level of sustainability.

\section{Conclusion}
\label{sec_conclusions}
Starting from the standard approaches to service stack decomposition, specifically layered stack decomposition and networked stack decomposition, a new noninvasive approach and mechanism to automation of cloud computing operation has been proposed. The proposed approach, called Northwise/Southwise Visibility/Transparency Pathways (NSVTPs), enables the northern components such as applications to directly communicate with southern components such as hardware without adding any additional requirement on the middle-layer components. The NSVTP provides this means of communication toward bypassing the opacity induced by the middle-layers and abstraction which have been well-accepted by-products of cloud computing. The core of the NSVTP approach is based on a concept called capsules that is then integrated with the resource identifiers in the form of appendices. Capsules enable exchange of blueprints, schemes, and formula between southern and northern components. These exchanges could be then leveraged toward a more optimal unified operation of all components. In a special use case related to dynamic frequency scaling and DVFS, it has been shown that the NSVTP could improve the `relative' energy performance with a further minimum of 60\% even when it is added to the highly-adaptive and optimized disaggregated hardware solutions.

\section*{Acknowledgment}
The authors thank the NSERC of Canada for their financial support under Grant CRDPJ 424371-11 and under the Canada Research Chair in Sustainable Smart Eco-Cloud (NSERC-950-229052), and also the MITACS of Canada.

\begingroup
\bibliographystyle{IEEEtran} \bibliography{imagep}
\endgroup

\setcounter{section}{0}
\renewcommand{\thesection}{Appendix \arabic{section}}
\section{Cloud and Virtualization}
\label{sec_Cloud_and_Virtualization}
There are well-established and agreed-on standards available for both cloud computing and virtualization. For example, we can refer to the ITU-T recommendation with respect to the cloud computing's overview and vocabulary \cite{ITUY3500201408I}. These standards, recommendation and guidelines are especially designed and used in the context of service request/provide interactions and transactions between cloud computing providers and users/customers. However, we think that the core of these technologies could be described at even a lower level. We acknowledge that such low-level description may not be feasible to be mapped to the provider/user interactions.

\subsection{Cloud}
\label{sec_cloud_definition}
Cloud Computing could be imagined as a plurality of computing operations on top of some `nameless' resources. Although pluralized resources could be used to create a state of nameless resources, it is not the only possible way. 
In other words, we could reach the core of the Cloud Computing approach, i.e., relaxation of the dedicated-resource constraint, without enforcing plurality to the resources.

In terms of all efforts related to Cloud Computing standards, it could be mentioned that their coverage and convergence has been well carried out which is a key requirement for handling business operations. However, by nature, this has introduced a 'black-box' perspective that in turn masks the actual internal technologies and mechanisms used in each solution that granted the title of Cloud Computing. For the purpose of having an alternative 'white-box' perspective, it would be helpful to call one of the possible classes of internal approaches the Cloud Approach. This specific class would cover all those solutions which make resource 'names' (IDs) impermanent in order to allow 'resources' to be seamlessly changed (swapped, added, or dropped) by the user or the cloud manager. This class would include many possible approaches including those that stand on top of the Virtualization technologies. The virtual resources generated by virtualization approaches are by nature ephemeral and therefore their names carry the same ephemerality feature.

\subsection{Virtualization}
\label{sec_virtualization_definition}
Virtualization could be seen as a way to make a resource plural. In other words, this technology provides a way to have isolated resource-sharing. This makes virtualization a good choice to separate and also to abstract the actual [physical] resources, which in turn makes virtualization a widely-used approach to enable cloud computing (with replacing the dedicated-resource constraint with resource pooling from a large set of short-living, ephemeral virtual resources). However, as mentioned in the previous section, it is possible to deliver a cloud computing solution without using virtualization. 

To provide a metaphoric example, let us consider the case of vehicles. A 'Vehicle' could be defined a thing that provides the transport `function' by relaxing the fixed-position constraints of some of its parts. In contrast, 'Wheel'-ization enables a disk object [resource] to have plural states (different angles) around its axis. Many vehicular types are delivered using the wheel technology. However, the question is that is it possible to have a vehicle without using wheels? And the answer in this metaphoric example is yes: For example, many construction machines, which are also vehicles, do not use wheels and instead they may use Continuous Tracks. However, the question would be then whether we can use these low-level definition of a vehicle in the market for provider/user interactions? The answer is simply no. Depending on jurisdiction, there are many standards and conditions that a vehicle should satisfy to be eligible to enter the market and the roads. For example, all vehicles should comply to Motor Vehicle Safety Regulations (C.R.C., c. 1038) in Canada. The some logic would hold for the cloud technology. However, the regulatory guidelines and standards should not be literally taken as the `boundary' of a technology by the developers: With introduction of any new instance or new feature to the cloud technology, there will be a period time of convergence needed to include that new instance/feature in the market regulations.

In the following section, we introduce a new feature in the cloud computing that could bring a higher level of performance and therefore greenness and sustainability to the operations if other influential factors are set properly. This feature, which we call the Northwise/Southwise Visibility/Transparency Pathways (NSVTPs), enables direct, real-time interaction among components of a cloud computing solution even if they are opaqued and blocked by the other middle-layer components. In order to put the NSVTP mechanisms in a proper context, we first define two major approaches to service decomposition in the cloud computing. 
Any effort to improve the overall sustainability and reduce resource consumption can be seen as an enhancement to cloud solutions. One category of such efforts can be modeled in the form modifying paths of operations into `Re-Cycle' {\em paths}. These paths could be then augmented by other alternatives such as Up-Cycle modifications. Generally, we could categorize these modified paths in a bigger class of {\em X-Cycle} modifications. It is worth mentioning that such enhancements have had also an indirect positive consequence in the role of ICT as an enabler of sustainability in the non-ICT operations \cite{Malmodin2015}. 

\section{Layered Stacks vs Networked Stacks: NFV-Native Architectures}
\label{sec_Layered_Stacks_vs_Networked_Stacks_NFVNative_Architectures}
Although the Layered (in either Single-Column or Multi-Column forms) approach to the design of computing {\em stacks} has been generally favored, new approaches such as that of the Networked Stacks, which will be discussed in section \ref{sec_Networked_Stacks}, have been also considered mainly because of their specific advantages especially in the context of the `function'-oriented architectures \cite{Farrahi2016a,Bogineni2016}. Below, we first review the definitions of the layered and networked stacks, and then we will explore how it would be possible to add the proposed {\em pathways} to these types of stacks in the next section.

\begin{figure*}[tbh!]
	\centering
	\begin{tabular}{@{}cc}
		\fbox{\includegraphics[height=2in]{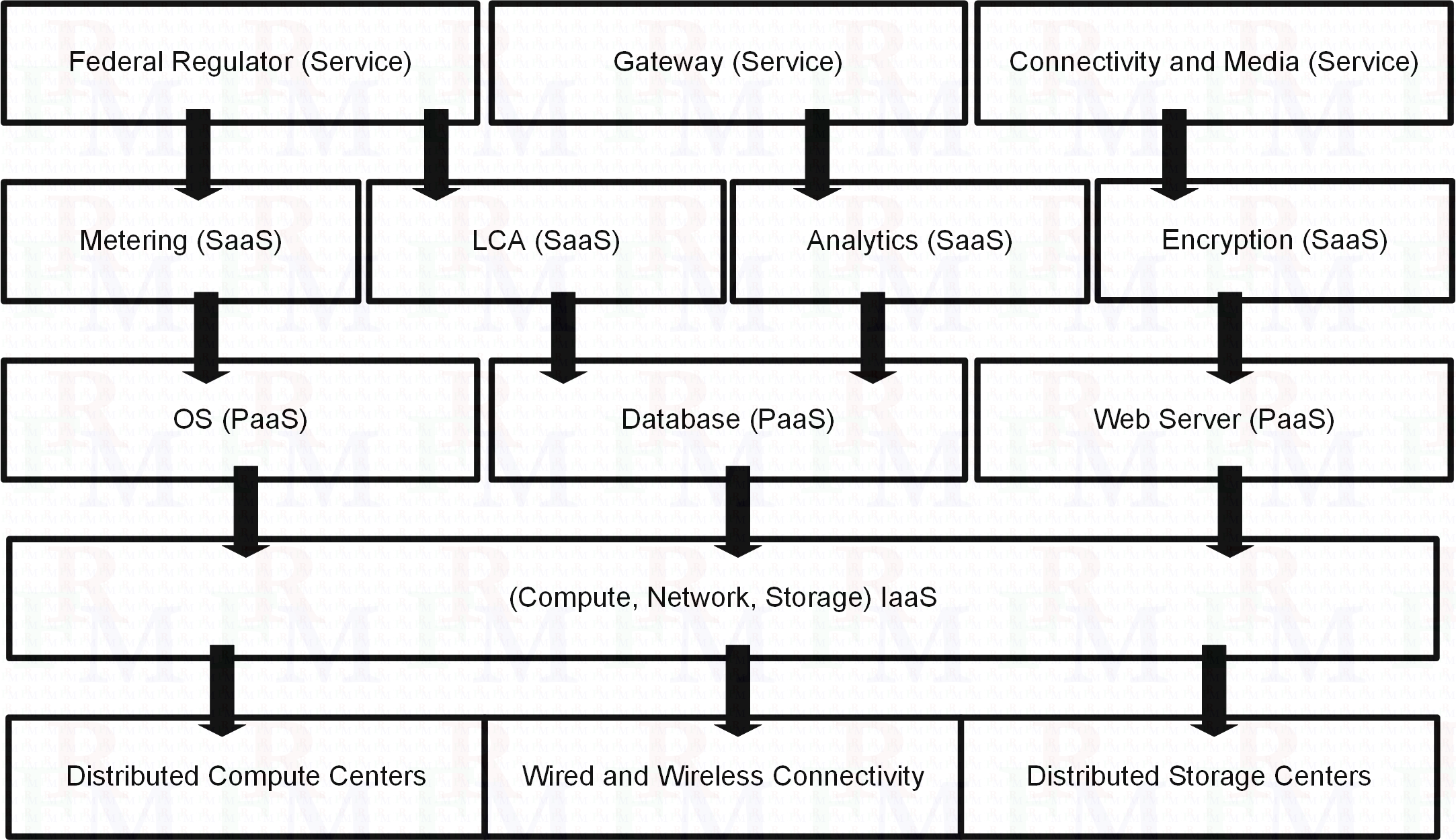}} &
		\fbox{\includegraphics[height=2in]{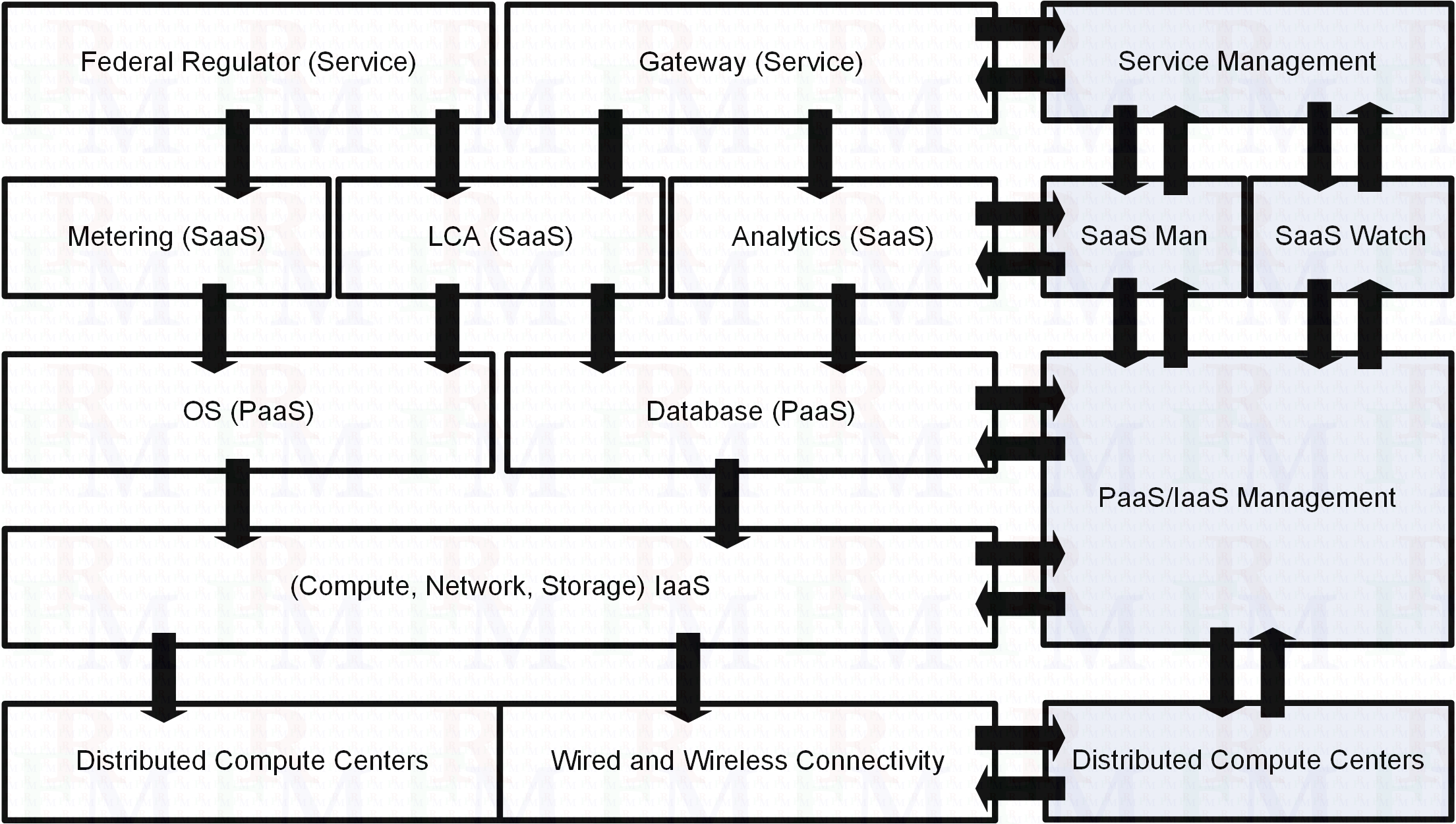}} \\
		(a) & (b) 
	\end{tabular}
	\caption{Layered Service Stack.
		a) A Single-Column Layered Stack.
		b) A Multi-Column Layered Stack.}
	\label{fig_Layered_Service_Stack_Decomposition1}
\end{figure*}

\subsection{Layered Stacks [Approach to Service Decomposition]}
\label{sec_Layered_Stacks} A Layered Service-Stack Decomposition is a realization of a possibly-complex service in terms of an `layer-indexed' set of a few simpler services/components. The simpler services are constrained to {\em only} interact with those other simple services that are their immediate neighbors in terms of the {\em layer} order. For example, a service with a layer {\em index} of the number 5 could only interact with services with the layer index numbers 4 and 6.\footnote{It is worth mentioning that there could be exceptions in which a layer extends across more {\em than} one layer, and therefore could potentially interact with components that are at two different but adjunct layer index.}\footnote{Please note that the layer `index' is not limited to be only {\em numeric}. For example, the set of \{Infrastructure, Platform, Software, Service\} could serve as the set of layer indices. However, we require the existence of an {\em order} among indices.} Any layered service decomposition could be visualized as a stack of simple services placed on top of each other according on their layer-order (for example, starting from the components with the {\em lowest} order index). In such a visualization, each simple component is considered as a service-layer component, and the whole picture is called a layered service-stack (see Figure \ref{fig_Layered_Service_Stack_Decomposition1}). 

\begin{figure*}[tbh!]
	\centering
	\begin{tabular}{@{}cc}
		\fbox{\includegraphics[height=1.9in]{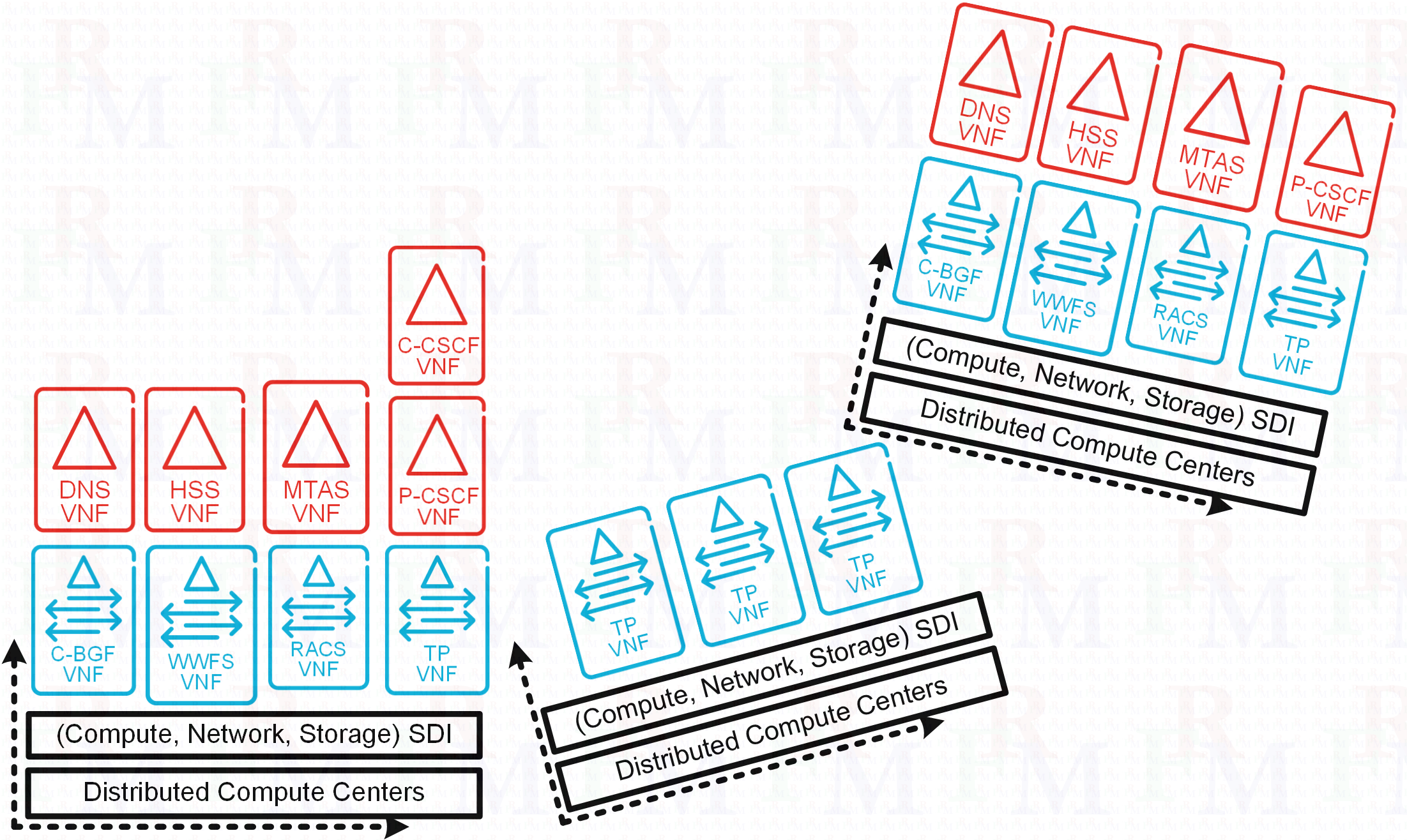}} & 
		\fbox{\includegraphics[height=1.9in]{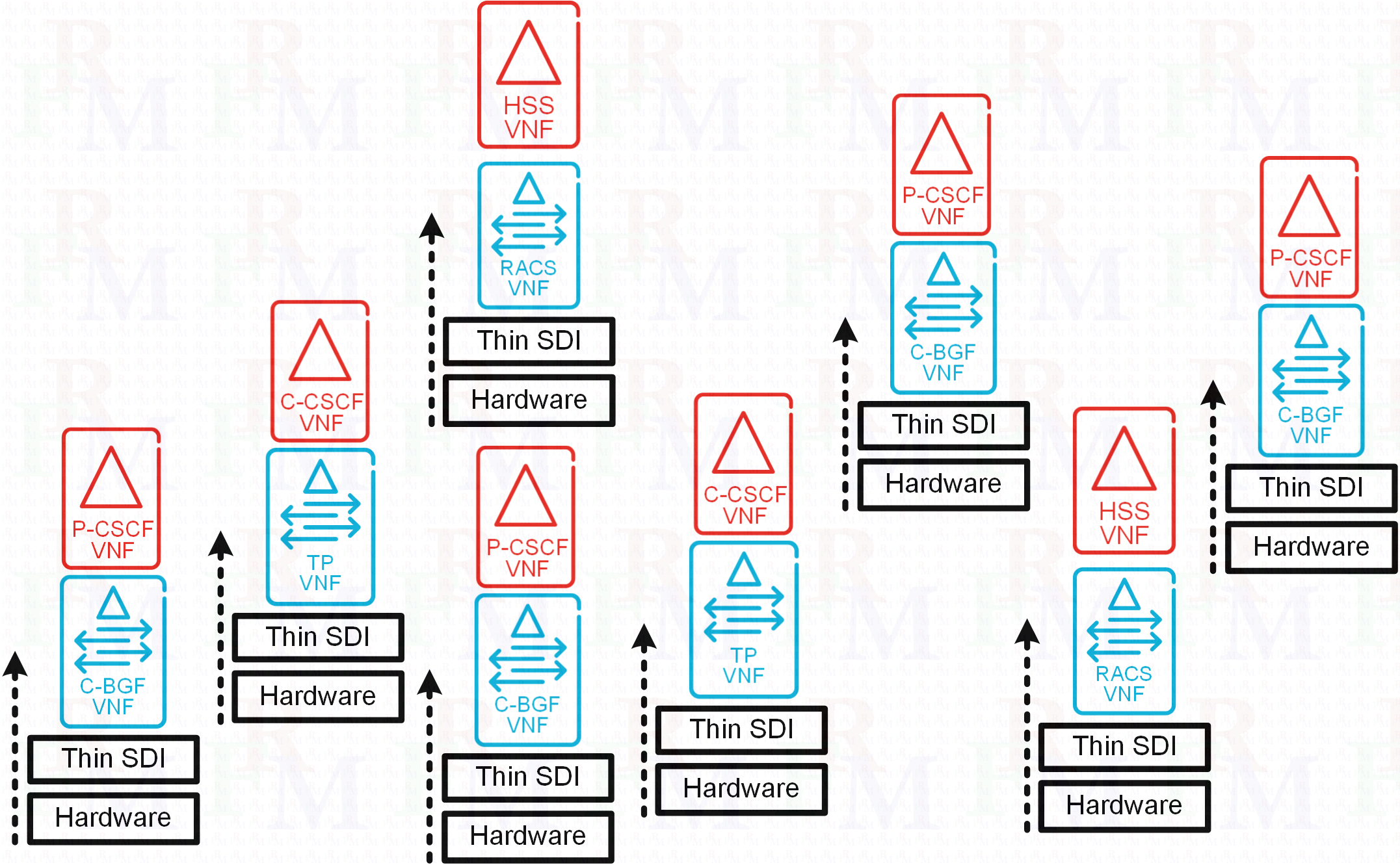}} \\
		(a) & (b) 
	\end{tabular}
	\caption{Networked Service Stack.
		a) Quasi-Linear networked Stack. The horizontal axis indicates the `location' and the vertical axis refers to the `functions'.
		b) Unconstrained Networked Stack. There is no significant role for the location within the small islands. However, the functions axis is always relevant.}
	\label{fig_Networkedd_Service_Stack_Decomposition1}
\end{figure*}

\subsection{Networked Stacks [Approach to Service Decomposition]}
\label{sec_Networked_Stacks}
In contrast to the Layered Service-Stack Decomposition approach (and also the Chained Service-Stack Decomposition approach \cite{Farrahi2016a} which is not mentioned here), in the networked Service-Stack Decomposition there is no constraint in terms of the order and also in terms of the number of egress connectors exiting from a component (see Figure \ref{fig_Networkedd_Service_Stack_Decomposition1}). In other words, any simple service could egress to more than one other simple services. A direct consequence of this relaxation of constraints would show itself as a challenge in the service orchestration, i.e., the `timing' of service requests among simple services. At the same time, networked stacks have a great potentials in translating complex services in the form of functional relations among a plurality of simple components. The function-oriented approach has been of great interest especially in the context of Network Function Virtualization (NFV) toward less over-provisioned, more efficient and performance, and therefore more sustainable ICT operations. It is worth mentioning that the networked Stacks could partially support layer-order in some cases. We call such stacks quasi-linear networked stacks. An example is shown in Figure \ref{fig_Networkedd_Service_Stack_Decomposition1}(a). In this figure, each cluster of Virtualized Network Functions (VNFs) stands on top of a thin layer of software defined infrastructure (SDI). The concept of pathways, which will be discussed in the next section for the layered stack decompositions, could directly applied within each `island' of VNFs. The generalization of the pathways approach to inter-island interactions, which will discussed in the future, is of great interest when the networked stack decompositions move outside the quasi-linear form. In the unconstrained form, shown for example in Figure \ref{fig_Networkedd_Service_Stack_Decomposition1}(b), there is no horizontal concept of location applicable within each island. However, still the vertical dimension along the functionalities is applicable and is used in the pathway mechanisms. To populate the VNFs in the figure, we used a few sample functions from the IP Multimedia Subsystem (IMS) solutions \cite{ETSITS123228Y2015,Bogineni2016}: Home Subscriber Service (HSS), Border Gateway Function (C-BGF), Resource and Admission Control Sub-System (RACS), Proxy-Call Session Control Function (P-CSCF), Serving-Call Session Control Function (S-CSCF), WebRTC\footnote{WebRTC: Web Real-Time Communication} Web Server Function (WWSF), and Teleport (TP). Here, we added the TP function which allows seamless function-calling among VNF islands.

\end{document}